\documentclass{article}
\usepackage[utf8]{inputenc}
\usepackage{graphicx}

\usepackage[margin=1in]{geometry}
\usepackage{titlesec}
\usepackage{enumitem}
\usepackage{hyperref}
\usepackage{url}

\title{Recommender Systems in Financial Trading: Using machine-based conviction analysis in an explainable AI investment framework}

\author{Alicia Vidler \\ \textit{School of Computer Science and Engineering} \\ \textit{UNSW, Sydney} \\
\textit{a.vidler@unsw.edu.au}}
\date{March 2022}

\begin{document}
\maketitle

\begin{abstract}

Traditionally, assets are selected for inclusion in a portfolio (long or short) by human analysts. Teams of human portfolio managers (PMs) seek to weigh and balance these securities using optimisation methods and other portfolio construction processes. Often, human PMs consider human analyst recommendations against the backdrop of the analyst's recommendation track record and the applicability of the analyst to the recommendation they provide. Many firms regularly ask analysts to provide a "conviction" level on their recommendations. In the eyes of PMs, understanding a human analyst's track record has typically come down to basic spread sheet tabulation or, at best, a "virtual portfolio" paper trading book to keep track of results of recommendations.

Analysts' conviction around their recommendations and their "paper trading" track record are two crucial workflow components between analysts and portfolio construction. Many human PMs may not even appreciate that they factor these data points into their decision-making logic. This chapter explores how Artificial Intelligence (AI) can be used to replicate these two steps and bridge the gap between AI data analytics and AI-based portfolio construction methods. This field of AI is referred to as Recommender Systems (RS). This chapter will further explore what metadata that RS systems functionally supply to downstream systems and their features.
\end{abstract}


\section{Introduction}

\textbf{Definition:} Recommender systems (RS) ‘are information filtering and decision supporting systems that present items in which the user is likely to be interested in a specific context.\footnote{N K Liu and KK Lee, ‘An intelligent business advisor system for stock investment’, Expert Systems, 14(3), 129–139, (1997).}

A Recommender system is a form of AI and is typically a standalone piece of software or computer code. Recommender systems (RS) are designed to "recommend" a course of action to the end decision maker, or more commonly, another piece of software that makes further analysis. RS's are typically helpful when dealing with a data problem of selecting from a universe of items that is many orders of magnitude larger than the number of items a user wishes to select. For example, if PM's theoretical investible universe covers some 1000 -- 5000 listed securities, a typical portfolio they manage may contain 100 or few stocks, a small percentage of the universe of all possible choices.

Recommnder systems were initially a niche application of a broader, earlier form of AI, Expert Systems, which came to prominence in the 1980s. The academic field proliferated following the publication of the "Handbook of Applied Expert Systems" in 1997. One of the first academic works by Lui and Lee of Hong Kong University in 1997 considered a system that was an "intelligent business advisor system for stock investment and was widely implemented and offered a list of features for analysing and picking stocks, based on user preferences which the authors noted needed to be supplied by the end investor. RS shot to prominence more broadly with the Netflix Prize, a competition with USD \$1m in prize money for a solution to produce the best collaborative filtering algorithm for Netflix, then an online DVD rental platform. The prize was finally won in 2009. Since then, many online consumer tools in areas such as music (Spotify), shopping (Amazon) and news media have continued to advance the field. Extensive academic work exists in the field of applied RS as early as the 2000s, with meta review studies being published as early as 2016 covering over 100 papers \cite{HE20169}.

\section{Common Types of RS AI and their mainstream adoption}
Four common subtypes of RS AI exist: collaborative filtering, content-based filtering, knowledge-based systems and case-based systems. Knowledge-based systems are the most applicable and widely used subtypes in asset selection and institutional investing.

\begin{description}[font=\normalfont\itshape, leftmargin=0cm]
    \item[Collaborative Filtering:] AI which, rather than using specific data's inherent or apparent features, a meta-model and user responses classify data. A basic example might be that the AI utilises public tweets or blog posts that mention the stock rather than using the listing location as a data field.
    
    \item[Content-based filtering] uses the inherent features of an item and incorporates past items selected by a user that are filtered based on the same inherent features. These typically have lower predictive powers in financial settings and rely upon multiple criteria decision analysis (MCDA) techniques. As the name suggests, users must extensively specify these systems and explain why they made previous stock choices.
    
    \item[Case-based reasoning] uses predefined "cases" stipulated by users. These are then applied to other time periods of evaluation.
    
    \item[Knowledge-based recommendation systems:] The most widely implemented form of RS in finance. These are complex systems that have collected and systematised expert humans' knowledge and human experts' decision-making processes. They often involve highly technical analysis components that seek insight from large data sets, e.g., trading price data of single stocks. Additionally, these systems use ranking and sorting methods to produce personalised recommendations based on user specifications, training data, historical versions of the same software and currently available market data.
\end{description}

In the past five years, there has been a significant increase in the number of available AI-based software packages and platforms capable of taking proprietary client data and forming bespoke recommendations. Many tools have long tried to incorporate other forms of AI into their recommendation engines including fuzzy logic methods, artificial neural networks (at the post recommendation analysis phase) and even methods such as support vector machines when trying to filter recommendations with the added usage of data such as news information. Machine Learning techniques can also be applied to the calibration methods within recommendation systems. The key message to take away here is that many AI methods can be applied sequentially or within a broader AI method. In the following section I present a fictional case study of a senior PM at a hedge fund (or investment firm) to illustrate the applicability of AI methods applied to individual asset classes.

\section{Features, details, and design}
To help illustrate a complex implementation, I will make use of a fictional case study. I will focus on how a firm of trading and investment professionals, utilise teams of internal analysts to develop strategies on individual assets and implement a RS. For ease of discussion, I will assume we are talking about an equity trader, and the asset class that they look at are Asian and European equities. This case study could apply to any listed security market.

\subsection{An example of a Senior PM at a hedge fund or investment firm and their journey with RSs}
A typical day would involve a PM either directly or in conjunction with a small team of 1 -5 analysts running analysis and internal algorithms on the securities in the universe that make up her trading mandate. For this example, where we use Asian and European equities, liquidity and market capitalisation constraints are crucial considerations, dictating our PM's starting universe of possible securities. As is most common, PMs will tend to focus on either specific sectors or regions. For many reasons related to capital constraints and regulation, they will most likely look at firms with market capitalisations above some minimum threshold. This clearly defined universe will have other consequences; it will dictate the data the PM can use in standard software and the risk limits they can use. Suppose our PM has the mandate to trade long and short. In that case, the availability of stock borrow for cover of short positions will also form the initial stock universe.

A team of analysts will typically run in-house scripts, code, spreadsheets, or even internal programs looking for patterns and previously identified trading strategies against the universe of securities. Some PMs will have this process automated internally; others who trade with a lower frequency will look to have human analysts build "high conviction" lists for proposed long and short trades. Except for the sophistication of the automation of these steps defined here, little in this workflow has changed in over two decades in a PM's day to day role.

In the early 2000s, many stock exchanges started to increase the availability of their market trading data and allow computers to link to their electronic order books directly and explicitly.  Those equity exchanges that were completely voice-driven moved to electronic trading screens and dealer interfaces.  Those largely screen-based exchanges invested heavily in the necessary hardware to support faster data access and trading capabilities.  What has resulted is a sea of market data.  By the late 2000s and into the 2010s, many firms had grown internal specialised teams of programmers and analysts.  These teams are highly skilled at fast data analysis and produced robust in-house technology infrastructure to repeatably and reliably automate the daily data analysis process.  However, the output of many of these firms that still maintained traditional PMs skills was to produce information for a human PM to digest, synthesise and ultimately use to make discretionary trading decisions.  A small but highly successful group of quantitative funds flourished from the mid-2000s onward and specialised in technology.  Firms traditionally focused on discretionary trading now have PMs who deal with an ever-increasing deluge of semi-processed market information each day. Gradually, as firms acknowledged the increasing size and challenge of garnering insight from partially processed data, a sort of “virtual” analyst evolved, which technically is a form of RS.  The financial market journey parallels many of the similar advances made in non-financial market products, such as consumer software used in online shopping or digital media around the same time.

As a human team of analysts would present their high conviction buy and sell recommendations to our hypothetical PM, people build RS to produce virtual analyst recommendations.  Systems began to develop organically from outsourced software providers as early as 2010 and are increasingly popular and well adopted today.  One key reason for the increasing adoption of RS is the advent of a field of AI called Explainable AI, or Causal AI.  This academic field focuses on what a RS and AI system need to show an end-user to elicit trust in their virtual recommendation. Many currently available software packages rely heavily on graphs and visualisation techniques to illustrate intermediate steps in the overall recommendation process.  Such methods appear to be users’ preferred format for gaining trust in otherwise black box AI systems in this field.   

Several firms are dedicated to cloud-based AI algorithms looking at stock price information.  Hedge funds and asset managers who use these cloud-based software packages can log in and run proprietary or generic AI trading strategies on single stocks or baskets.  Many firms will do a similar thing across data from other industries regarding medical clinical trials and online sales data from consumer goods platforms.  These cloud-based AI firms looking at stock information require the PM to design their algorithm within their framework, import their data, or use existing options.  The output of these systems is what differentiates this group of RS from other forms of software: they provide relative recommendations, come with some supporting information about the decision and usually some heuristic interpretation of the recommendation. 

A key feature is RS by design rarely, if ever, recommended a course of action such as “buy” or “sell”.  So, what does a RS output look like? A list of buys ranked by some pre-agreed analysis method from highest conviction to lowest conviction.  Some systems will enrich the idea of “conviction” to help explain the system’s idea of a human notion of confidence.  For example, a sample track record may be applied to each conviction on each stock that made it into a ranked suggestion list of “buys” – where the conviction of the algorithm producing the recommendation may show a “hit ratio” of that algorithm on that stock over the past 100 days. You can begin to see that many practical forms of conviction analysis can be done here.  For example, perhaps the PM might want to know how the algorithm has performed on a given stock over the past 10 days, 50 days, and 100 days because that aligns with the PM’s macro view on the current markets.  Similarly, a PM with a long-term trading strategy may only be interested in a conviction analysis of 100 or more days.  Some may look for time period analysis that follows a geometric progression or even a much-beloved Fibonacci number sequence. Just like single stock technical trading styles favour different technical indicators for different market environments, commonly available RS have many varieties of preloaded and embedded AI-based algorithms – though they tend to be knowledge-based AI systems.  

Returning to our PM, they may seek to use an in-house implementation of a RS, uploaded and calibrated to her specific macro or market views, and within the risk management framework constraints of her trading mandate.  The PM might request a ranked list of high conviction stock trading suggestions and high conviction short trading suggestions to meet the market-neutral nature of her overall portfolio constraint (noting that long and short portfolios may not guarantee any in and of themselves market neutrality).  they might even have sector or country limits that they incorporated into her RS to receive a ranked and sorted recommendation list that has factored these into their recommendations.  They may also use one of several commercially available third-party cloud-based software providers rather than an in-house software solution.  Once received, our PM will typically follow one of two courses of action; the first is to review these recommendations manually (using human intuition) or, the second, to import the recommendations (along with associated metadata) into another technology system. Whilst external software platforms for recommendations are a new invention, older styles of in-house software that seamlessly passed data from a RS into other downstream internal systems are still far more common.  

\subsection{Dealing with a RSs set of recommendations: what next?}
Turning again to our PM, assuming they are one of the more sophisticated PMs, looking to make use of every piece of data at their disposal, now a different analysis process occurs.  The research comes into play in various fields of old and new AI.  Older fields of RS or expert systems can form a basis of rigid “if this, then that” rules that run over these recommendations to refine a list of trades long and short.  Systems like these have been covered in academic literature from the 1970s.  Several large, prestigious, and long-established hedge funds imply they use these methods in addition to high-frequency trading firms (though with slightly different applications).

More recently, however, as AI research has increased and explainable AI, causal AI and even “ethical AI” have grown in importance and relevance, newer methods of dealing with the output from algorithmic trading recommendation systems have emerged.  These fields have all been spurred along by increasing human adoption of AI in people’s personal lives (think AI-enabled home devices, online grocery shopping apps and ride-sharing apps that seem to know where you are travelling to before you even type in an address).  In addition, increasingly, regulatory, and legal responsibilities are being placed on AI.  Namely, human end-users of AI interim recommendations and decisions need to be comfortable with those decisions as they, in many jurisdictions, carry the legal responsibility for those trading decisions.  Hence, humans need to understand the recommendations made and essentially own the risk that taking decisions based on them brings. 

So, what might an example be?  Returning to our busy PM, her ranked, sorted list of long and short high conviction stock recommendations would be loaded up into different pieces of software.  This downstream software will typically fall into three broad groups – 

\begin{itemize}
\item Risk management – pre-trade analysis 
\item Portfolio management – pre-trade analysis
\item Pure alpha analysis – PM specific further analysis

\end{itemize}

As experience in modern portfolio management will attest, today’s money management firms have a plethora of internal software – everything from books and records systems to order management systems and at least one compliance and one risk management software package.  Many firms may use up to 15 different software packages in a given day, all with specific niche applications that remain crucial to any investment process’s efficient and compliant running.  A straightforward example of this is the need in many jurisdictions to source stock borrow for a short position before placing a trade that will take a specific trading account into a short position.  Firstly, finding the required stock borrow can bring market challenges (such as telegraphing to other firms the desire to short a holding), not to mention the administrative challenge of making sure it is internally assigned to the correct PM’s trading account. These are all burdensome processes, without which our PM may find herself non-compliant in certain jurisdictions.  What is more crucial to appreciate here is that the PM will need more information than simply a list of ranked buy and sell recommendations to put into “downstream” systems.  These additional details are commonly referred to as “metadata” and are crucial for the correct technical functioning of independent technology software such as below;

\subsubsection{Risk management system analysis}
The PM might analyse the impact of recommendations on the risk characteristics of her current portfolio.  Any sound risk system will incorporate “what if” scenario analysis.  Perhaps the PM may have a more manual process for incorporating a dummy portfolio to look at its risk features and how it would have performed historically.  More than the possible returns of the portfolio, the use of risk systems would allow the PM to check that by incorporating the recommendations, no strict risk limits and mandate restrictions would be broken.  More technically advanced trading firms, or those readers with a strong high frequency trading (HFT) background, might perform this analysis in a systematic automated process through database manipulation and other scripting/programmatic methods.  Essentially the same outcome is sort: how the recommended ranked and sorted list of buy and sell ideas would affect the current holdings of the PM.  

In terms of “metadata” needed to perform these stages, the ranked and sorted list of recommendations really would need to pass appropriate identification tickers for securities and some way to turn that into either a proposed number of shares to buy or sell, or more as a percentage of the ideal portfolio the PM is trying to target.  Special care needs to be taken here in the choice of stock price used to calculate either, as prices need to be synchronous with the risk system data and the analysis periods used in the initial recommendation systems.  This detail is often missed and can produce a sense of “forsight” of a model that can bias more automated backtesting processes.  

\subsubsection{Portfolio management – pre-trade analysis}
Here, the PM would use similar data fed to a risk system to evaluate the performance attributes of the proposed buys and sells.  Depending on the sophistication of the PM and the software used, this may include shadow profit and loss (PNL) forecasting and even parsimonious methods for shadow PNL forecast attribution.  Whilst outside the scope of this chapter, both outcomes will ideally need to know more about how a buy or sell recommendation was ranked in the conviction recommendations.  

This brings us to a crucial aspect of Explainable AI in particular – how does one piece of AI explain to a human (or even a second piece of software that a human uses to visualise it) why it ranked its recommendations the way it did?  For those unfamiliar with this type of analysis, it's important to understand how basic AI programs operate when performing linear program analysis (i.e., simple "if this, then that" logic). The AI software can quickly explain how it arrived at its decisions because it can essentially retrace the path it took through all the conditional statements ("ifs" and "else's") to reach a particular conclusion or ranking.

It can even output all the various maths calculations it used to come up with some mark or score, and then the way it thought about ranking them (e.g., biggest to smallest or smallest to biggest, or even removing outliers then taking some ordered list).  However, the methodologies here are almost endless; the point is that in older AI forms, the methods are linear and easy to describe in a tree diagram or “list of steps”.

We run into more difficulty with more advances and modern forms of nonlinear AI; that is to say, they do not follow a clear linear path from input to output, with the same steps repeated for the same inputs every time.  For those of you not familiar with nonlinear AI methods, you may have heard of things such as Machine Learning, neural networks, random forest, deep learning algorithms and fuzzy logic – to name a few.  In their most straightforward format, in an attempt to be smarter than a simple list of “if this then that” commands, AI has developed to think more like humans naturally do (which is associative) and to think in entirely unstructured unintuitive ways (e.g., neural networks, random connections tested over large amounts of history).  Newer methods can be compelling as they introduce an element of uncertainty and artificially “lateral” thinking.  This “lateral” thinking component is a strength and weakness, making such AI hard to explain and potentially untrustworthy to a human user.  The following section will deal with practical examples of these issues.

\subsubsection{Decision Support Systems - alpha analysis and PM specific further analysis}
The third use case for recommendation systems output is for our PM to upload the ranked and sorted lists into an alpha analysis program that provides decision support services.  Alpha analysis programs can be as simple as a spreadsheet or as complex as several commercially available AI supercomputer frameworks (e.g., IBM Watson or TensorFlow).  

Her team of analysts could even produce the program in Python or C++. The location or type of program matters little; what is critical here is the reason for analysis and the methods utilised to look at the recommendations from “many angles”. The straightforward idea is to try and take the recommendations and correlate them independently to other macro factors such as a currency rate or interest rate curve, looking to identify a relationship with independent variables.  This process itself could be done as a form of machine learning.  However, most experienced PMs would be concerned about taking a series of recommendations and arbitrarily testing them for correlations to other variables where little or no use of the actual variable was used in coming up with a recommendation – this would be fertile ground for discovering spurious correlations.  More likely, as we saw in our discussion on the Portfolio management software, the PM would want the recommendation software to pass along metadata that allowed more extensive analysis in a testing environment controlled by the PM.  Extensive back testing is some of the tests they might run beyond those already completed in a risk and portfolio setting. So how would a nonlinear AI explain its own decisions?  Welcome to the world of explainable AI.  One critical method is a counterfactual explanation, another more practical process is paper trading records.

\subsection{How to explain RS outputs, how to explain AI-led recommendations}
The field of explainable AI is a huge and growing one.  In practice, there remains little consensus on the ultimate “perfect” explanation.  Instead, practically speaking, people look to things that have tended to replicate existing frameworks of explanation in other domains.  For example, when a senior PM looks at her team of PMs and analysts, they might naturally recall their profit and loss records, their trading decisions around specific events, and when her analysts and PMs got decisions very right and very wrong. They might also recall the over-exuberance of analysts who presented investment decision cases with high degrees of certainty, only to be wrong-footed by the market a few days later as some unexpected implicit assumption no longer held.  Depending on her level of experience and that of her team, they may recall live examples of how people traded, and decisions made over particularly turbulent market events such as the 2008 Global Financial Crisis or Greek bond default and the COVID pandemic.  These examples can be easily replicated with AI by asking the RS to produce a paper track record of its various recommendations over time.  These track records created within the recommended system (or outside of it by comparing many days’ worth of recommendations) provide Machine Learning algorithms to data-mine and learn and adapt decision-making layers. 

Many forms of AI are primarily done without the interference of human users, others require training data and others still like to verify aspects of their processes at various points in time with factual human-verified inputs.  These latter classes can be thought of as reinforcement and supervised learning methods.  One simple outcome of all this analysis might be that our PM’s internal alpha system can gauge how best to interpret the recommendations from the initial RS program.  Such interpretations might include, but would not be limited to, thinking about how good the RS sorting and ranking methods are, and if, for example, the top recommendations might end up putting the PM into already crowded trades or positions with existing very high short interest.  These ‘insights’ would require much data and Machine Learning calibration and may or may not be stable through time.  Understanding these insights and their weaknesses will become our PM and her team’s core skill and chief value-add.  

\section{Big Data tools and RS utilisation}
The process described features increasing amounts of data.  The very concept of time series analysis is such that by virtue of the process described above existing for six months, there is an additional day of history at each additional day.  Added to this is the issue that RS that impart or directly rely on some training data or data needed for lookback periods, are notorious for using sizeable historical data sets.  Furthermore, depending on the investment horizon and risk management framework of the investment manager in question, modelling intraday data for investments may be necessary.  If the RS focuses on investment idea generation, trading implementation, and best execution analysis, significantly more market data will be needed.  Any analysis of order execution or intraday risk management intervention would typically require very granular trading information, possibly down to individual tick by tick analysis of the securities in question.  

All this adds up to a considerable amount of data.  Different components of the analysis process utilise different aspects of data and, potentially, different time granularity of data.  Returning to a PM and her team, setting aside the technical challenges of such systems, humans have a strong preference of being about to verify data inputs visually.  Many advanced tools now exist for data visualisation.  What is perhaps novel in an RS is the need to visualise raw data inputs (e.g., price trends) and intermediate variable states and intermediate decisions.  As discussed above, not all forms of AI used by RSs can provide transparency (visual or otherwise) that will satisfy a human or replicate current processes.  At this point, visualisation tools will aim to interpret the various mathematical methods of explanation that nonlinear AI models will need. 

As any good programmer will know, one should never re-invent the wheel when commoditised code exisits that can be quickly and efficiently sourced external to a firm (e.g. repos on Github).  The past decade has seen an explosion in outsourced data analysis firms, many focusing on finance and robustly providing the exact commoditised analysis processes that I mention here.  Much like the early 1990s saw the advent of sophisticated statistical analysis programs, and even the extensive expansion of capability within Microsoft Excel, from around 2012, there has been a similar trend in analysis software from portfolio return attribution analysis through to outsourced Machine Learning tools and database AI tools. Many of these early outsourced analysis platforms were high-quality but initially struggled with client adoption.  Next, I will detail some of the reasons for this and how overcoming such obstacles has directly affected how RS work.

\section{Considerations and complications of outsourced analysis platforms.}

We have already seen the evolving need for more robust and industrial-strength analysis platforms to understand, interpret, and evaluate AI decisions, particularly those from RS in portfolio management settings.

Outsourced software, which has grown in popularity and capability, has always provided significant value to those firms willing to use them.  However, their adoption has only recently gained pace, with many of these software firms announcing prominent venture capital and investor support alone in the past 12 months.  

There are four key categories in terms of practical considerations a PM or their firm might make before allowing the use of such outsourced software.

\begin{itemize}
\item \textbf{Intellectual Property (IP): Protection and Creation}.
Any use of a technical system tends to remain the user's IP; however, it is the fundamental nature of AI that some degree of learning from one users' needs is used to enhance the overall product (for other users).  The only way to avoid this would be to have a standalone in-house system  - but then again, your firm might not benefit from general product development long term.  Practically speaking, this can be addressed by segmenting the work done in one disparate system to another. 
\item \textbf{Robustness and support}. 
A key feature of external software is the size of the team and the technical advantage of specialised IT support.  Even small analytic firms often have teams of engineers and programmers who outnumber even a large group at a single financial institution.  The increasingly specialised nature of analytic technology means that external software can be a key leverage point for smaller PMs and investment managers. 
\item  \textbf{Integration and quality control}. 
There is always a point at which the end-user no longer has visibility around the software when using external software.  Also, many internal software packages at large investment firms remain obscured to the end-user.  In the case of Machine Learning, there is a component of proprietary implementation of ubiquitous processes that complicates the picture.  External software providers need to keep some aspects of their code proprietary.  Hence, integration and quality control for the PM or end-user becomes essential.  Internal IT staff to a PM’s firm may change their role to monitor outsourced software for changes rather than maintaining in-house software packages.  Again, such monitoring tools exist and are pretty standard in cyber security software.     
\end{itemize}

\section{Trends and future development}
I have outlined here many directions that AI in RS are taking, such as explainability and methods of garnering the “trust” of human users. In researching this chapter two such trends were most notable; the method systems are using to garner trust, and the people who are most frequently using such systems.  

Human users of RS software today appear to be demanding more and more visualisation of key data components and intermediate analysis.  This is leading many software providers to develop highly specialised and visually appealing user interfaces.  The actual data that appears to be displayed, at the request of the user, is not necessarily the most important data used by the RS, but rather reflects the interest of users into specific data.  

Furthermore, external RS software today seems targeted to the large number of discretionary managers and “traditional” asset investors.  This contrasts with many forms of specialised AI software which rely on users having advanced understanding of AI.  RS software targeted at discretionary managers is potentially going to “bridge the gap” between the vast majority of non-technically trained investment staff and the increasing power of AI applied to investing.  It remains to be seen if software firms are able to provide user experiences that can be acceptable to their clients without any loss of power of the AI behind their investment recommendations. 

\section{Summary}
As the adoption of AI, systems, analytics, and broadly based recommendation engines increases, many people look to quantify the uplift in returns that such tools can provide.  Whilst looking to see what an AI trading system would have traded, and its resulting profit and loss record over a period is helpful; the bigger picture is fast becoming which firms can afford not to use any AI systems at all.  Such has been the adoption by investors and consumers alike to the power and potential of AI that many firms now appear to feel they need to say they are using some forms of it to remain relevant and current to the landscape.  This chapter has shown the difference between merely purchasing a license to some AI-enabled data platform and the actual integration of recommendation-based AI systems in a portfolio management context.

\section{Acknowledgments}
I would like to express my appreciation to all those who provided me the possibility to complete this report.  A special note of thanks to Dr Middleweek, (University of Technology Sydney), for her careful review and Professor Walsh (Scientia Professor of AI, University of New South Wales) for his thoughtful suggestions.

\nocite{*}


\bibliographystyle{plain}
\bibliography{bib}

\end{document}